\documentclass[aps,twocolumn,showpacs]{revtex4}
\usepackage{amssymb}
\usepackage{amsfonts}
\usepackage{booktabs}
\usepackage{graphicx}
\begin{document}


\title{Transport properties and asymmetric scattering in Ba$_{1-x}$K$_x$Fe$_2$As$_2$ single crystals compared to the electron doped counterparts Ba(Fe$_{1-x}$Co$_{x}$)$_{2}$As$_{2}$}

\author{Bing Shen$^{1}$, Huan Yang$^{2}$, Zhao-Sheng Wang$^{1}$, Fei Han$^{1}$, Bin Zeng$^{1}$, Lei Shan$^{1}$, Cong Ren$^{1}$ and Hai-Hu
Wen$^{2}$$^{\star}$} \affiliation{$^{1}$Institute of Physics and
Beijing National Laboratory for Condensed Matter Physics, Chinese
Academy of Sciences, P.O. Box 603, Beijing 100190, China}
\affiliation{$^{2}$National Laboratory for Solid State
Microstructures, Department of Physics, Nanjing University, 210093
Nanjing, China}

\begin{abstract}
Resistivity, Hall effect and magnetoresistance have been
investigated systematically on single crystals of
Ba$_{1-x}$K$_x$Fe$_2$As$_2$ ranging from undoped to optimally doped
regions. A systematic evolution of the quasiparticle scattering has
been observed. It is found that the resistivity in the normal state
of Ba$_{1-x}$K$_x$Fe$_2$As$_2$ is insensitive to the potassium
doping concentration, which is very different from the electron
doped counterpart Ba(Fe$_{1-x}$Co$_{x}$)$_{2}$As$_{2}$, where the
resistivity at 300 K reduces to half value of the undoped one when
the system is optimally doped. In stark contrast, the Hall
coefficient R$_H$ changes suddenly from a negative value in the
undoped sample to a positive one with slight K-doping, and it keeps
lowering with further doping. We interpret this dichotomy due to the
asymmetric scattering rate in the hole and the electron pockets with
much higher mobility of the latter. The magnetoresistivity shows
also a non-monotonic doping dependence indicating an anomalous
feature at about 80 K to 100 K, even in the optimally doped sample,
which is associated with a possible pseudogap feature. In the low
temperature region, it seems that the resistivity has the similar
values when superconductivity sets in disregarding the different
T$_c$ values, which indicates a novel mechanism of the
superconductivity. A linear feature of resistivity $\rho_{ab}$ vs.
$T$ was observed just above $T_c$ for the optimally doped sample,
suggesting a quantum criticality.

\end{abstract}


\maketitle

\newpage

\section{Introduction}
The discovery of iron-based superconductors~\cite{Kamihara} has
triggered great interests in the field of condensed matter physics.
A lot of theoretical and experimental works suggest complicated
Fermi surfaces and unconventional pairing
mechanism\cite{I.I.Mazin,Kuroki,LeeDH,HDing,Gordon,Cren,BZeng}.
There is a common issue between the iron pnictides and the cuprates,
that in both systems the superconductivity is in the vicinity of the
antiferromagnetic (AF) order, leading to a very similar phase
diagram. However, the feature of the undoped parent phase is
actually quite different. The cuprate may be categorized as the
so-called Mott insulator, while the iron pnictide is a poor metal.
The phase diagram has thus been a focus of intense research in order
to pursue an understanding of the relationship between the AFM and
the superconducting (SC) states, and intimately the superconducting
mechanism\cite{JZhao,LFang,NNiBaFeCo,Johrendt,Hchen}. For the
electron doped 122 family, such as Ba(Fe$_{1-x}$Co$_x$)$_2$As$_2$, a
lot of works both in experiment and theory illustrate the systemic
evolution of the transport properties and electronic structure, all
indicate the importance of the multiband effect\cite{R.Prozorov, F.
Rullier-Albenque,ImaiNMR,J. J. Tu, LFang, M. A. Tanatar}. It was
presumably believed that in the underdoped regime, the AF and the SC
phase compete for the density of states along the Fermi surfaces. In
the overdoped regime superconductivity suffers from a suppression of
the spin fluctuations\cite{ImaiNMR}, and probably the loss of the
Fermi surface nesting. These two effects lead to an asymmetric
superconducting dome, which is somewhat different from the case in
the cuprate superconductors. By analyzing the transport \cite{LFang}
and optical data\cite{J. J. Tu,WangNL1} in
Ba(Fe$_{1-x}$Co$_x$)$_2$As$_2$, it has been consistently claimed
that the mobility of the electron band are much higher than that in
the hole band. It is curious to know in a hole-doped sample, would
the mobility disparity survive, disappear, or change sign? In this
paper we report the systematic studies on resistivity, Hall effect
and magnetoresistivity on selected hole doped
Ba$_{1-x}$K$_x$Fe$_2$As$_2$ single crystals. The detailed
investigations and analysis suggest the asymmetric quasiparticle
scattering in the hole and electron band, with still a much larger
mobility in the electron band. Meanwhile we present the evidence
that the electron doping induced by substituting the Fe sites with
Co results in a great impurity scattering, but without breaking too
much Cooper pairs. These Fe-sites doping may generate the impurities
which can only scatter the electrons with small momentum transfer,
with the inter-pocket pairing un-intact.

\begin{figure}
\includegraphics[scale=0.8]{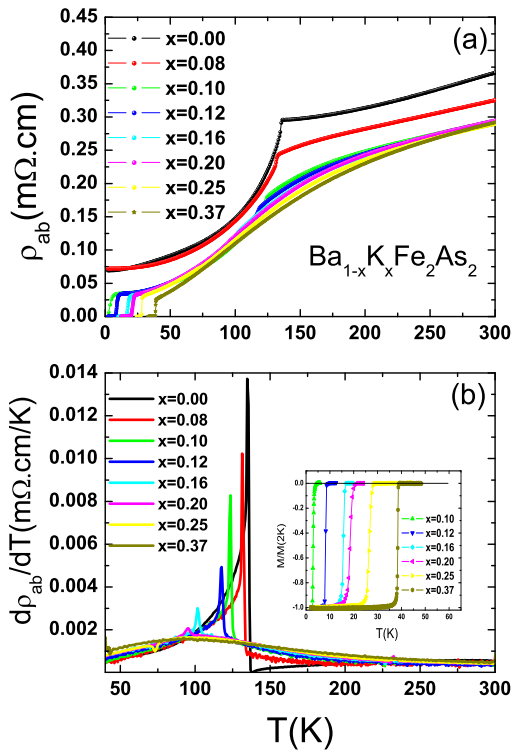}
\caption{(Color online) (a) Temperature dependence of resistivity of
Ba$_{1-x}$K$_x$Fe$_2$As$_2$ (x=$0 \sim$ 0.39) single crystals. The
AF / structural transition is shifted to lower temperatures and
becomes invisible with further potassium doping. The resistivity
anomaly in normal state can not be explicitly resolved when the
doping level is 0.25 with T$_c$ = 29 K, and beyond. (b) Temperature
dependence of derivative of the resistivity d$\rho_{ab}$/dT. The
peak in d$\rho_{ab}$/dT associated with the AF / structural
transition is suppressed with doping and disappears at x = 0.25.
Inset: Zero field cooling magnetization of six superconductive
samples.} \label{fig:fig1}
\end{figure}

\section{Experimental}
The Ba$_{1-x}$K$_x$Fe$_2$As$_2$ single crystals were grown by
self-flux method using FeAs as the flux, detailed procedures of
synthesizing the samples are similar to the previous
reports\cite{IRFisher,HQLuo,Lshan}. The crystal structure and
chemical composition were checked by X-ray diffraction and energy
dispersive X-ray microanalysis. For the transport measurements, all
the samples were cut into rectangular shape and the standard six
electric probes were made by silver paste. The electronic transport
measurements were carried out in a Physical Properties Measurement
System (PPMS, Quantum Design) with the temperature down to 2 K and
magnetic field up to 9 T. The superconducting transition temperature
of the samples were determined by the 50\% of the normal state
resistivity. The in-plane longitudinal and the Hall resistance were
measured by either sweeping the magnetic field at a fixed
temperature or sweeping the temperature at a fixed magnetic field.
Both sets of data coincide with each other.

\section{Results and Discussion}
\subsection{Temperature and doping dependence of resistivity}
Fig.1(a) shows the temperature dependence of resistivity for
Ba$_{1-x}$K$_x$Fe$_2$As$_2$ single crystals with doping levels
ranging from undoped parent phase to optimally doped compounds. It
is clear that the potassium doping makes the system evolve from AF
(with a resistivity anomaly) to superconductive. The sharp drop of
resistivity at about 138 K due to the AF / structural transition can
be observed in the parent phase. With doping holes, the resistivity
anomaly is suppressed and shifts to lower temperatures, which is
agreeable with the peak in the derivative of the resistivity
d$\rho$/dT shown in Fig.1(b). When the doping level x reaches 0.25,
the resistivity anomaly disappears and a little pit can be observed
in d$\rho$/dT where the superconducting transition temperature is 29
K. With further doping, the T$_c$ reaches a maximum value at 39 K.
For the electron-doped Fe-based 122 family, the magnetic and
structural transition are slightly separated\cite{IRFisher} leading
to two close peaks on the d$\rho$/dT vs. T curves. The resistivity
anomaly exhibits two kinks which are corresponding to the two peaks
in d$\rho$/dT for the underdoped samples. While for hole-doped
BaFe$_2$As$_2$, some experiments show that the AF and structure
transition occur at the same temperature in underdoped region
\cite{Hchen,Johrendt,Johrendt2}. In our measurements, we observed
only one single sharp peak in d$\rho$/dT vs. T curves for underdoped
Ba$_{1-x}$K$_x$Fe$_2$As$_2$ which is agreeable with previous
results.

Here we would like to emphasize several contrasting issues by
comparing the resistivity curves in electron doped
Ba(Fe$_{1-x}$Co$_{x}$)$_{2}$As$_{2}$ (Co-122)\cite{LFang} and hole
doped Ba$_{1-x}$K$_x$Fe$_2$As$_2$ (K-122) samples (Fig.1). Firstly,
in the Co-122 samples, the resistivity at 300 K reduces to half
value of undoped sample when the system is optimally doped
\cite{LFang}, this is actually not the case in the K-122. One can
see from Fig.1(a) that, the resistivity at 300 K drops only about
20-30$\%$ when the doping is getting to the maximum value. What is
more interesting is that the resistivity value when the
superconductivity sets in is quite close to each other although the
T$_c$ changes from 10 K (x=0.10) to 29 K (x=0.25).  This may suggest
the resistivity is mainly dominated by the electron band, which is
weakly influenced by the hole doping. The interest triggered by this
observation is two fold: the threshold for the occurrence of the
superconductivity is governed by the residual resistivity at T$_c$,
while the T$_c$ value is determined by the hole concentration,
perhaps by how strong the suppression to the antiferromagnetic phase
is. It is not clear at this moment what leads to this strange
behavior, but clearly it indicates a novel mechanism of the
superconductivity. Secondly, the resistivity exhibits an up-rising
step at the AF/structural transition in the electron doped Co-122,
while in K-122, this transition exhibits always as a drop of
resistivity at T$_{AF}$ and it is smeared up gradually with more
doping. Thirdly, the RRR, namely the ratio between the room
temperature resistivity and the residual resistivity (just above
T$_c$) is about 2.4 in Co-122,\cite{LFang} indicating a strong
impurity scattering. But it seems such a strong scattering does not
block the superconductivity. Based on the picture of pairing through
inter-pocket scattering,\cite{I.I.Mazin,Kuroki} the non-magnetic
impurities may be detrimental to the supercnductivity if they induce
the inter-pocket scattering. In this sense, the impurities here may
induce the scattering only with small momentum transfer, for
example, intra-pocket scattering. In the optimally hole doped
samples, the RRR can get to 14, indicating a weak impurity
scattering. At high temperatures, the $\rho_{ab}$-T curve shows a
bending down feature for the hole doped samples. In the conventional
single band metal, the bending down of resistivity was interpreted
as the approaching to the Ioffe-Regel limit, which is corresponding
to the case that the mean free path induced by the phonon scattering
is comparable to the atomic lattice constant. This seems not the
case here, since the electron doped sample Co-122 has the same
structure and similar phonon spectrum, but the bending down feature
of resistivity has not been observed up to 300 K.

Above mentioned behavior of the resistivity can be qualitatively
understood by the two band scenario with asymmetric scattering rate
in the hole and the electron pockets. According to the simple two
band model, the conductivity can be written as
$\sigma=\sum_{i}\sigma_i$, with $\sigma_i=n_ie^2\tau_i/m_i$ the
conductivity, $n_i$ the charge carrier density, $\tau_i$ the
relaxation time, $m_i$ the mass of the $i$-th band ($i$ = $e$ or $h$
for the electron and hole band, respectively). Therefore the
resistivity can be described as

\begin{equation}
\rho=\frac{m_em_h}{e^2(n_e\tau_em_h+n_h\tau_hm_e)}.
\end{equation}

It is known that the parent phase has identical area of electron and
hole Fermi surfaces in the non-magnetic phase, therefore we can
assume an identical charge carrier densities $n_e$ and $n_h$ for the
two bands ($n_e\approx n_h\approx n_0$). Considering m$_h$$ > $m$_e$
as revealed by ARPES\cite{DingH} and specific heat\cite{MuG}, and
assuming that $\tau_e > \tau_h$, the conductivity is thus dominated
by electron band. With the electron doping, the term $n_e\tau_em_h$
is getting much larger than $n_h\tau_hm_e$ which reduces the
resistivity further. To the optimally doping at about x = 0.08,
$n_e$ has increased a lot, perhaps doubled, this reduces the
resistivity to almost its half. In the case of K-122, the situation
is different. If still adopting the relation of
$n_e\tau_em_h>>n_h\tau_hm_e$, doping holes will decrease $n_e$ but
increase $n_h$, in this case the resistivity should increase,
instead of decrease. Actually doping holes will on one hand decrease
$n_e$ and increase $n_h$, but more important is to lower down the
$m_h$ and $1/\tau_h$, in this case, the resistivity will be
determined by a balance between these quantities and shows a weak
doping dependence. A quantitative understanding would require a
detailed doping dependence of $n_i$, $\tau_i$, $m_i$ ($i$ = $e$,
$h$). This is out of the scheme of what we can get from a simple
resistivity measurement and analysis. Some calculations indicate
that the conductivity for electrons grows strongly upon electron
doping, while the hole conductivity varies weakly compared to that
of the electrons\cite{Hirschfeld}. Thus the resistivity of
hole-doped Ba$_{1-x}$K$_x$Fe$_2$As$_2$ at 300 K changes less than
that for electron-doped Ba(Fe$_{1-x}$Co$_{x}$)$_{2}$As$_{2}$. The
multi-band effect that one band is strongly coupled and relatively
clean, while the other band is weakly coupled and characterized by
much stronger impurity scattering will cause anomalous T-dependence
of the in-plane resistivity: the curve is convex with the tendency
to saturate at high temperature\cite{Golubov}.

\begin{figure}
\includegraphics[scale=0.7]{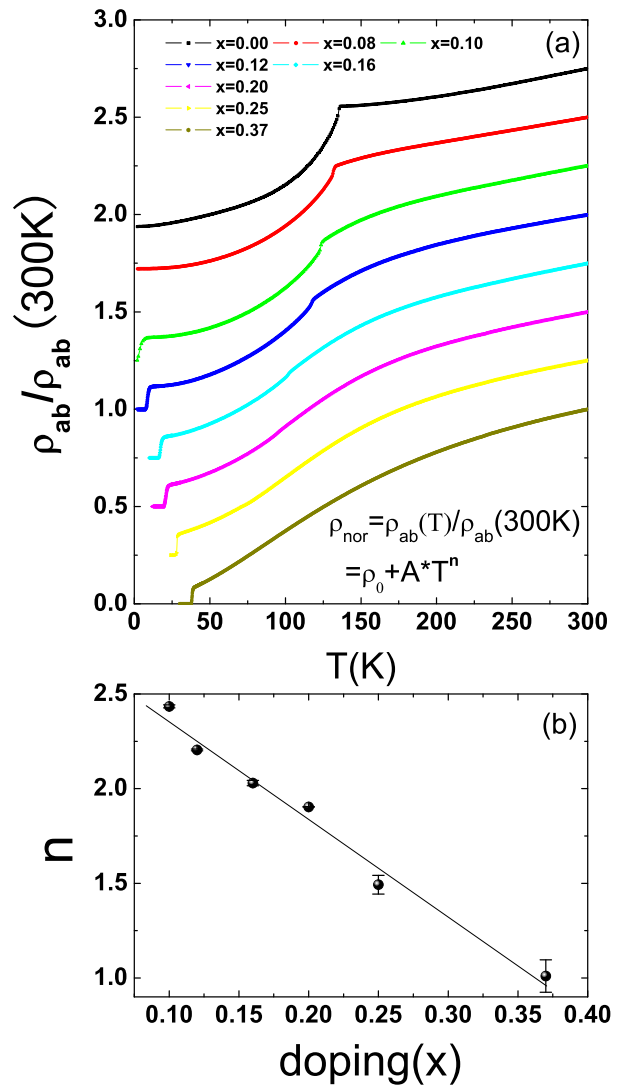}
\caption{ (Color online) (a) Temperature dependence of the
normalized in-plane resistivity ($\rho_{ab}/\rho_{ab}(300 K)$) of
Ba$_{1-x}$K$_x$Fe$_2$As$_2$. The data sets are offset vertically by
0.25 for clarity. (b) The fitting parameter n for six doped
superconducting samples (see text).}\label{fig:fig2}
\end{figure}

\begin{figure}
\includegraphics[scale=0.65]{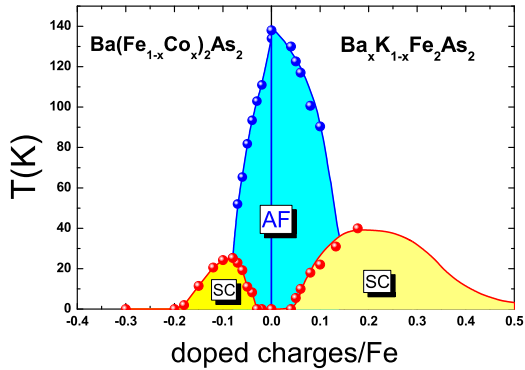}
\caption{(Color online) The phase diagram of
Ba$_{1-x}$K$_x$Fe$_2$As$_2$ and
Ba(Fe$_{1-x}$Co$_{x}$)$_{2}$As$_{2}$. No splitting between the
structural and AFM transitions has been observed in the hole doped
samples. The curve of T$_c$ vs. doping in wide hole doping region
(solid line) was taken from the Ref.\cite{Johrendt}. The data in the
electron doped regions were adopted from Ref.\cite{LFang}. }
\label{fig:fig3}
\end{figure}

\subsection{Trace of possible quantum critical point at the optimal doping}
In Fig.2(a) we present temperature dependence of the normalized
$\rho$ by the room temperature resistivity $\rho_{ab}$(300K). A
quick glance at the data can immediately see that the $\rho$-T curve
in the low temperature region changes from a non-linear to a linear
behavior towards optimal doping. In order to know precisely the
evolution of the resistivity with doping, we fit the data in the low
temperature region by the equation
\begin{equation}
\rho_{nor}=\rho(T)/\rho(300 K)=\rho_0+A\times T^n
\end{equation}
with three fitting parameters $\rho_0$, A and n for each curve. Due
to the saturation in the resistivity at high temperatures and the
anomalies of resistivity induced by the AF/structure transition, we
fit the data below the AF/structure transition (for x=0.25 and
below). For the optimally doped sample, the fitting was done with
the data between 40 and 120 K. The resulted fitting parameters are
presented in Table I and the exponent n is shown in Fig.2(b). The
evolving from a power law with exponent n = 2.3 to a linear
temperature dependence can be easily observed in Fig.2(b), which may
indicate the crossover from a Fermi liquid behavior to non-Fermi
liquid when the quantum critical point is approached. It was
previously pointed out that the exponent n in metals near an AFM
quantum critical point (QCP) may be sensitive to disorder\cite{A.
Rosch}. While in K-122 the impurity scattering is quite weak, this
can be corroborated by the negligible $\rho_0$ value at the optimal
doping. At the optimal doping, the T-linear resistivity in the low
temperature region may suggest a quantum critical point. Similar
behaviors have been observed in
Sr$_{1-x}$K$_x$Fe$_2$As$_2$\cite{chu}, BaFe$_2$As$_{2-x}$P$_x$
\cite{Y. Nakai}, Ba(Fe$_{1-x}$Co$_{x}$)$_{2}$As$_{2}$ \cite{ImaiNMR}
etc.. It has been pointed out that the quantum fluctuation becomes
very strong when the Neel temperature of the AFM order becomes zero.
It is this strong quantum fluctuation that heavily couple to the
itinerant electrons and modifies the transport property. Although it
was argued that this linear feature may be reconstructed with a
residual term and a T$^2$ term in the optimally doped
Ba(Fe$_{1-x}$Co$_{x}$)$_{2}$As$_{2}$ system,\cite{Barisic}, the
systematic evolution shown in our present study can rule out this
possibility.

For the cuprate superconductors, the antiferromagnetic order of the
magnetic moments of the Cu$^{2+}$ is completely suppressed before
superconductivity sets in. They do not coexist at any point of the
T$_c$(p) (p: doped hole number) phase diagram (exception was
suggested in the Bi-2201 system). In contrast, the coexistence of
the AFM and the superconductivity can be observed in underdoped
region of the dome of K-122 \cite{Hchen,ImaiNMR,Y. Zhang,F.
Massee,Johrendt2}. Fig.3 shows the phase diagram of
Ba$_{1-x}$K$_x$Fe$_2$As$_2$ and
Ba(Fe$_{1-x}$Co$_{x}$)$_{2}$As$_{2}$. Although there are some
reports claiming that magnetically ordered phases and SC state are
probably mesoscopically/microscopically separated \cite{T. Goko,J.
T. Park,A. A. Aczel}, most of the studies on K or Co doped samples
are in favor for the coexistence of magnetic order and
superconductivity and have consistently ruled out the presence of
phase separation\cite{ImaiNMR,F. Massee,D. Johrendt}. The very small
residual specific coefficient $\gamma_0$ in the optimally doped
Ba$_{0.6}$K$_{0.4}$Fe$_2$As$_2$ also strongly suggest the absence of
macroscopic phase separation, since otherwise one should be able to
see a large residual term of specific heat. Therefore we argue that
the QCP occurs near the optimally doped samples where the AFM order
vanishes at about zero K. In the specific heat measurements, we also
found that the mass enhancement $m^*/m$ goes up quickly when the
optimally doping point is approached.\cite{Mug} To confirm the
existence of quantum critical point and the coexistence of magnetic
order and superconductivity need certainly extra investigations
using other local probes.

\begin{table}[!htbp]
 \caption{\label{tab:test}Fit Parameters}
 \begin{tabular}{|c|c|c|c|c|c}
 \hline \hline
  \toprule    x &  $\rho_0$ & A (10$^{-5}$) & n \\
\hline
  \midrule
   0.10& 0.117$\pm$0.00038 &0.54$\pm$0.0139 & 2.33$\pm$0.00324\\
   \hline
   0.12& 0.114$\pm$0.00025 &1$\pm$0.026 & 2.20$\pm$0.00482\\
   \hline
   0.15& 0.096$\pm$0.00102 & 3$\pm$0.187 &2.02$\pm$0.0139\\
   \hline
   0.20& 0.091$\pm$0.00007& 5$\pm$0.037&1.90$\pm$0.0017\\
   \hline
   0.25& 0.057$\pm$0.00263 &32$\pm$7 &1.49$\pm$0.04921\\
   \hline
   0.39& -0.005$\pm$0.00349 &359$\pm$57 &1.01$\pm$0.1566\\
   \hline \hline
  \bottomrule
 \end{tabular}
\end{table}

\begin{figure}
\includegraphics[scale=0.85]{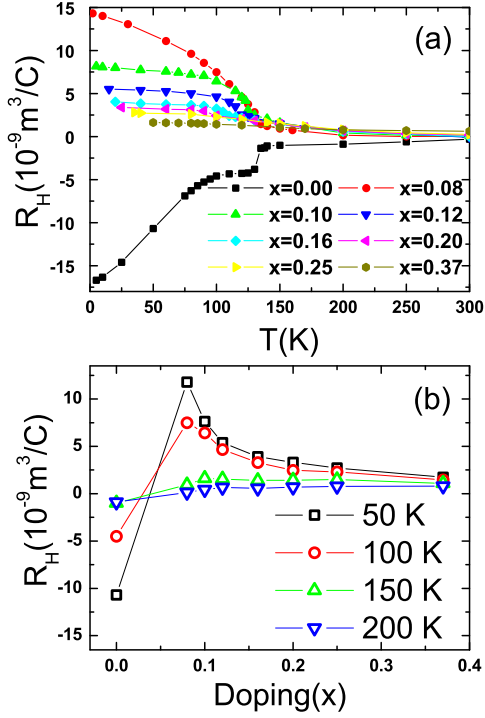}
\caption{(Color online) (a) The temperature dependence of Hall
coefficient of Ba$_{1-x}$K$_x$Fe$_2$As$_2$ ( x = 0 $\sim$ 0.37). The
Hall coefficient R$_H$ changes suddenly from a negative value in the
undoped sample to a positive one with slight K-doping. The
AF/structure transition can be found in the underdoping dome (0.08
$\leq$ x $\leq$ 0.25) as a onset of the rising of the Hall
coefficient R$_H$. For the optimal doped sample, R$_H$ varies less
pronounced with the temperature. (b) The doping dependence of Hall
coefficient at different temperatures.}\label{fig:fig4}
\end{figure}

\begin{figure}
\includegraphics[scale=0.7]{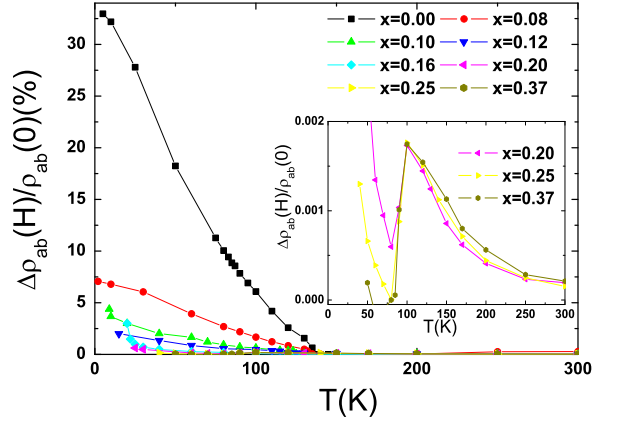}
\caption{(Color online) Temperature dependence of magnetoresistivity
for Ba$_{1-x}$K$_x$Fe$_2$As$_2$ (x=0$\sim$0.37) measured at 9 T.
Inset: magnetoresistivity of the samples (x=0.2, 0.25, 0.37). An
anomaly at around 80-100 K can be easily observed.} \label{fig:fig5}
\end{figure}

\subsection{Temperature and doping dependence of Hall coefficient}
The Hall coefficient R$_H$ from undoped BaFe$_2$As$_2$ to optimally
doped K-122 are presented in Fig.4(a) and the systematic evolution
can be observed. The Hall coefficient R$_H$ changes suddenly from a
negative value in the undoped sample to a positive one with slight
K-doping, and it keeps lowering with further doping in low
temperature region. For each doping level, the sudden increase of
the Hall coefficient is corresponding to the AF/structure transition
which is agreeable with the resistivity anomaly for underdoped K-122
(x = 0.1$\sim$0.25). Above the AF/structure transition temperature
the Hall coefficient varies weakly. The general formula for the Hall
coefficient in the Boltzmann approximation reads
\begin{equation}
R_{H}=\sum\frac{\sigma_i^2}{en_i}/(\sum\sigma_i)^2
\end{equation}

For fully compensated semimetals within the two-band model, Eq.3
reduces to
\begin{equation}
R_{H}=\frac{n_{h}\mu_{h}^{2}-n_{e}\mu_{e}^{2}}{e(n_{e}\mu_{e}+n_{h}\mu_{h})^{2}}.
\end{equation}
By definition, undoped samples are compensated, that is, n$_h$ =
n$_e$ = n$_0$. The  Eq. 4 can be written as
\begin{equation}
R_{H}=n_{0}^{-1}\frac{\mu_{h}-\mu_{e}}{\mu_{h}+\mu_{e}}=n_{0}^{-1}\frac{\sigma_{h}-\sigma_{e}}{\sigma_{h}+\sigma_{e}},
\end{equation}
where $\mu_i=\sigma_i/n_i=\tau_i/m_i$ ($i=e,h$) is the mobility. If
$\mu_{e}>>\mu_{h}$, then $R_{H}\approx1/en_{e}$, the transport is
dominated by the electron band\cite{LFang}. However, with the
potassium doping, the hole pocket increases in size instantly and
the electron pocket contracts. At the hole doping with an $x$ $\sim$
$n_{0}$, the Hall coefficient changes sign. With further doping, the
hole-doped systems have a presence of $\gamma$ pocket near
($\pi,\pi$). In addition to ($\pi$,0) scattering between $\alpha$
and $\beta$ sheets, new phase space for scattering opens
up.\cite{Hirschfeld}. The asymmetric scattering rate in the hole and
the electron pockets play an important role on R$_H$. Fig.4 shows
the doping dependence of Hall coefficient at 50 K, 100 K, 150 K, 200
K. In low temperature region, R$_H$ changes sign and reaches a large
value with little potassium doping. With further doping, R$_H$
decreases gradually. In high temperature region the R$_H$ varies
very little, which is agreeable with the recent
calculation\cite{Hirschfeld}. In the electron doped 122 system, the
transport property is dominated by electron. In compensated case,
the results can be explained by remarkable different mobilities of
hole and electrons\cite{LFang}. For hole doped K-122, the asymmetric
scattering rate in the hole and the electron pockets still holds,
but the relative ratio between $\tau_e/m_e$ and $\tau_h/m_h$ may
change a little bit, namely $\tau_h/m_h$ will get enhanced. This is
especially necessary to interpret the dropping down of resistance at
high temperatures when the slight holes are doped into the system,
as shown in Fig.1.

\subsection{Magnetoresistance}

The temperature dependence of magnetoresistivity for nine samples
measured at a magnetic field of 9 T are presented in Fig.5. The data
shows also a non-monotonic doping dependence and a sudden increase
below AF/structure transition, which is associated very well with
the anomaly found in resistivity and Hall effect. In undoped sample,
the large magnetoresistivity with a magnitude of about 35\% (at
about 9 T) has been found in low temperature region. With increasing
doping, the magnetoresistivity decreases instantly. It remains
unclear yet what causes this large magnetoresistance within the AF
phase. There are two main explanations: (1) The magnetic field will
break down the antiferromagnetic order to some extent and lead to
stronger spin fluctuations and thus larger scattering to itinerant
electrons; (2) A magnetic field will induce a stronger localization
leading to an enhanced resistivity.

In addition to this strong magnetoresistance in the
antiferromagnetic state, an anomalous feature at about 80 to 100 K
can be observed even in the optimally doped sample in which the AF
state dose not exist at all. One can see this in the inset of Fig.5.
The magnetoresistance rises up gradually when the temperature is
lowered down, but it drops suddenly at about 100 K and reaches
almost zero (for the optimally doped sample), then it rises up again
in the lower temperature region and smoothly connected to the
magnetoresistance induced by the vortex motion in the mixed state.
This anomaly at about 100 K may be associated with a possible
pseudogap feature due to some unknown reasons. This is consistent
with the recent observation in the c-axis resistive measurements
where a maximum of $\rho_c$ is observed.\cite{M. A. Tanatar} A high
temperature pseudogap was also claimed very recently from the
optical conductivity measurements.\cite{WangNL} Further experimental
and theoretical investigations are needed to clarify this point.

\section{conclusions}
In summary, we investigated resistivity, Hall effect and
magnetoresistance systematically on single crystals of
Ba$_{1-x}$K$_x$Fe$_2$As$_2$ ranging from undoped to optimally doped
samples. The resistivity in the normal state of
Ba$_{1-x}$K$_x$Fe$_2$As$_2$ is insensitive to the potassium doping
compared to the electron doped Ba(Fe$_{1-x}$Co$_{x}$)$_{2}$As$_{2}$
samples.  The Hall coefficient R$_H$ changes suddenly from a
negative value in the undoped sample to a positive one with slight
K-doping, and it keeps lowering with further doping. This
contrasting behavior is interpreted as due to the asymmetric
scattering between the electron and hole bands with the much larger
mobility in the former. An anomalous feature of magnetoresistivity
has been observed at about 80 to 100 K and may be associated with a
possible pseudogap feature. A linear feature of resistivity $\rho$
vs. $T$ was observed just above $T_c$ for the optimally doped
sample, which suggests a quantum criticality.

Acknowledgement: We acknowledge the fruitful discussions with I. I.
Mazin, M. A. Tanatar and P. Hirschfeld. This work is supported by the Natural
Science Foundation of China, the Ministry of Science and Technology
of China (973 project No: 2011CBA001002), and the Ministry of
Education of China (985 project).

 $^{\star}$ hhwen@nju.edu.cn

\end{document}